\documentclass[reqno]{amsart}
\usepackage{color}
\usepackage{eucal}
\usepackage{accents}

\newtheorem{theorem}{Theorem}[section]
\newtheorem{lemma}{Lemma}[section]
\newtheorem{proposition}{Proposition}[section]
\newtheorem{corollary}{Corollary}[section]

\theoremstyle{definition}
\newtheorem{definition}{Definition}[section]
\theoremstyle{remark}
\newtheorem{remark}{Remark}[section]

\newcommand{\xt}{\tilde x}
\newcommand{\xtt}{\tilde{\tilde x}}
\newcommand{\xh}{\hat x}
\newcommand{\xhh}{\hat{\hat x}}
\newcommand{\xth}{\hat{\tilde x}}
\newcommand{\xu}{\overline x}

\newcommand{\xtu}{\overline{\tilde x}}
\newcommand{\xhu}{\overline{\hat x}}
\newcommand{\ft}{\tilde f}

\newcommand{\fh}{\hat f}

\newcommand{\gt}{\tilde g}
\newcommand{\gtt}{\tilde{\tilde g}}
\newcommand{\gh}{\hat g}
\newcommand{\ghh}{\hat{\hat g}}
\newcommand{\gth}{\hat{\tilde g}}
\newcommand{\ght}{\tilde{\hat g}}
\newcommand{\bphi}{\boldsymbol \phi}
\newcommand{\bH}{{\bf H}}
\newcommand{\ovarphi}{\overline \varphi}
\newcommand{\opsi}{\overline \psi}

\def\XXint#1#2#3{{\setbox0=\hbox{$#1{#2#3}{\int}$}
     \vcenter{\hbox{$#2#3$}}\kern-.5\wd0}}

\numberwithin{equation}{section}
\begin{document}
\title{An Inverse Scattering Transform for the Lattice Potential KdV Equation}
\author{Samuel Butler, Nalini Joshi}
\address{School of Mathematics and Statistics F07, The University of Sydney, NSW 2006 Australia}
\email{s.butler@sydney.edu.au, nalini.joshi@sydney.edu.au}
\begin{abstract}
The lattice potential Korteweg-de Vries equation (LKdV) is a partial difference equation in two independent variables, which possesses many properties that are analogous to those of the celebrated Korteweg-de Vries equation. These include discrete soliton solutions, B\"acklund transformations and an associated linear problem, called a Lax pair, for which it provides the compatibility condition. In this paper, we solve the initial value problem for the LKdV equation through a discrete implementation of the inverse scattering transform method applied to the Lax pair. The initial value used for the LKdV equation is assumed to be real and decaying to zero as the absolute value of the discrete spatial variable approaches large values. An interesting feature of our approach is the solution of a discrete Gel'fand-Levitan equation. Moreover, we provide a complete characterization of reflectionless potentials and show that this leads to the Cauchy matrix form of $N$-soliton solutions.
\end{abstract}
\maketitle
\section{Introduction}

The lattice potential KdV equation 
\begin{equation}\label{eq:kdv}
Q_{pq}(x,\xt,\xh,\xth)=(\xth-x)(\xt-\xh)-p^2+q^2=0,
\end{equation}
where $x=x(m,n)$, $\xt=x(m+1,n)$, $\xh=x(m,n+1)$, $\xth=x(m+1,n+1)$ and $p$ and $q$ are complex lattice parameters, is an example of a nonlinear integrable lattice equation. It appears as the permutability condition for B\"acklund transformations of the KdV partial differential equation (see e.g. \cite{as:81}), transforms to the potential KdV partial differential equation under a particular continuum limit (see e.g. \cite{nc:95}), and has now been studied as an integrable lattice equation in its own right. \eqref{eq:kdv} possesses a 3D consistency property, a notion which has been studied in \cite{nw:01}, \cite{n:02}, \cite{bs:02} and \cite{abs:03}. In \cite{n:02} this property was shown to be equivalent to the existence of a Lax pair. In 2003 \eqref{eq:kdv} appeared as (H1) in the exhaustive list of integrable lattice equations in \cite{abs:03}, as a representative of a particular family of integrable lattice equations. One-soliton and two-soliton solutions are derived in \cite{av:04} by starting with a linear seed solution of \eqref{eq:kdv} and applying the B\"acklund transformation of the potential KdV partial differential equation. Furthermore in \cite{ohti:93} and \cite{nah:09} the authors give a determinant form for an $N$-soliton solution. In this paper we solve the initial value problem for \eqref{eq:kdv}, for initial profiles satisfying
\begin{align}\label{eq:IC1}
\sum_{m=-\infty}^{+\infty}|x_{m+2,0}-x_{m,0}-2p|(1+|m|)&<\infty,\\
\label{eq:IC2}
x_{m+2,0}-x_{m,0}&>0
\end{align}
along $n=0$. The $m$ and $n$ variables are related to the $\;\tilde{}\;$ and $\;\hat{}\;$ shifts by $\xt=x_{m+1,n}$ $\xh=x_{m,n+1}$. While it may appear that equation \eqref{eq:kdv} only depends on the single parameter $p^2-q^2$, it is natural to maintain the dependence on both $p$ and $q$ because each is associated with a different B\"acklund transformation, and one is therefore able to isolate the effect of each individual transformation. The initial conditions \eqref{eq:IC1} and \eqref{eq:IC2} are posed along the axis $n=0$, which is analogous to the continuous case, in which the initial condition is posed on $t=0$. This is the reason why the parameter $q$ associated with the $n$-direction does not appear here. The summability condition \eqref{eq:IC1} is a direct analogue of the integrability condition placed on the initial condition in the continuous case, and imposes the condition that the asymptotic behaviour of $x_{m+2,0}-x_{m,0}-2p$ decay faster than $m^{-2}$ as $|m|\rightarrow\infty$. Analogously to the continuous case, one finds for soliton solutions that this decay is exponential. 

\subsection{Background}
To our knowledge the first studies into the discrete inverse scattering transform date back to Case and Kac \cite{ck:73} and Case \cite{c:73}. These authors considered a direct discretisation of the time-independent Schr\"odinger equation and were led to an eigenvalue problem $A\phi=\lambda\phi$ where $A$ is a tridiagonal matrix. This was solved as an initial value problem for the half-line $n\ge0$. The inverse problem was posed on the unit circle in the complex plane of the spectral variable, and using the orthogonal polynomials that arose from the spectral distribution of $A$ the solution was obtained by deriving a discrete Gel'fand-Levitan integral equation. Flaschka \cite{f:74} showed how this procedure could be applied to solutions of the Toda lattice. The author considered linear difference equations in which the coefficients depended on the Hamiltonian of the lattice. These coefficients were assumed to depend smoothly on time and thus the discrete spectral data evolved according to a continuous evolution equation. The eigenvalues were shown to be constants of motion for all time, and the solution of the inverse problem was given as a solution of a discrete Gel'fand Levitan integral equation.

More recently Boiti et al \cite{bpps:01} considered an ``exact"\footnote{``Exact" meaning that the discretization arises directly from applying Darboux transformations to the Schr\"odinger equation} discretisation of the Schr\"odinger equation. They were led to a different discrete problem to \cite{ck:73}, \cite{c:73} and \cite{f:74}, one which had been studied earlier by Shabat \cite{s:99}, and coupled this with a two-parameter differential-difference time evolution equation, reminiscent of that given in the continuous inverse scattering transform in \cite{as:81}. Rather than following the Gel'fand-Levitan method for the inverse problem, the solution was given in terms of the time-dependent Jost solutions in the expansion of the spectral variable. These authors also considered a discretization of time, and postulated a similar evolution equation as in the continuous-time case. They found that this particular discretization led to a higher-order version of \eqref{eq:kdv}. 

This spectral problem in \cite{bpps:01} was later considered by Shabat \cite{s:02} as a dual problem to the spectral problem for the continuous Schr\"odinger equation. Using this duality property the author was able to use results of the continuous inverse scattering transform to obtain qualitative estimates for the discrete spectral problem. Levi and Petrera \cite{lp:07} obtained this spectral problem as one of the Lax equations for \eqref{eq:kdv} and used it to solve the inverse scattering problem for \eqref{eq:kdv}. Their discrete ``time" evolution was obtained using the second Lax equation and was different to that presented in \cite{bpps:01}, however the inverse problem was again given in terms of the Jost solutions and thus was a generalisation of that given in \cite{bpps:01}.  

In 2002 Ruijsenaars \cite{r:02} considered the same linear discrete problem as \cite{bpps:01}, coupled with a different parametrisation of the associated differential-difference time evolution equation. This equation depended on two independent variables $x$ and $t$ and contained iterates in $x$ and derivatives in $t$.  Ruijsenaars assumed analyticity in $x$, and considered the discrete problem as an eigenvalue problem for an analytic difference operator. This was preceded by a study of a class of reflectionless analytic difference operators that yielded soliton solutions. The inverse problem relies on a Hilbert transform, and the soliton solutions are shown to converge to regular KdV solitons under suitable scaling limits.  

In contrast the work presented here deals with the solution of the initial-value problem for \eqref{eq:kdv} as a function of two discrete independent variables. Thus both the linear problem and the scattering data evolve according to difference equations. We do assume continuity in the lattice parameters which allows the inverse problem to be solved by a Riemann-Hilbert approach, and we show that solutions of \eqref{eq:kdv} are given in terms of solutions to a discrete Gel'fand-Levitan integral equation.

\subsection{Outline of Results}

In this paper we solve the initial value problem for the LKdV \eqref{eq:kdv} rigorously through the inverse scattering transform method. Following \cite{n:02} we obtain a Lax pair for \eqref{eq:kdv} and derive the governing linear equation which agrees with that in \cite{lp:07}. In Sections \eqref{sec:direct}, \eqref{sec:Jost} and \eqref{sec:ab} the direct scattering procedure is carried out, in which the analyticity properties of the Jost functions in the plane of the spectral variable are proved rigorously, obtaining more precise bounds than those given in \cite{bpps:01}. We also give a sufficient condition for the poles of the transmission coefficient to be simple. In \cite{bpps:01} and \cite{lp:07} this property was assumed without proof, however this is not true in general for an arbitrary potential as is seen in the given counterexample. In Section \eqref{sec:n} the discrete ``time" evolution of the transmission and reflection coefficients is then derived, and the inverse problem is treated in Section \eqref{sec:inverse}. Rather than give the solution in terms of the ``time"-dependent Jost solutions, as was done in \cite{bpps:01} and \cite{lp:07}, we emulate the Gel'fand-levitan procedure for the continuous case and derive a discrete Gel'fand-Levitan integral equation
\[
K(m,L)+B(L)+\sum_{r=-\infty}^{m}K(m,r)(B(r-m+L)+B(r-m+L-1))=0,
\]
where $B$ is dependent on the scattering data. The solution $K(m,L)$ of this linear equation is then related to the solution of \eqref{eq:kdv} by
\[
x_{m+2,n}-x_{m,n}=2p\left[\frac{1+K(m+2,m+2)}{1+K(m+1,m+1)}\right].
\]
In Section \eqref{sec:12sol} we consider one- and two-soliton examples and finally in Section \eqref{sec:Nsol} we show that any reflectionless potential that satisfies the required summability and positivity conditions gives rise to an $N$-soliton solution identical to that found by applying B\"acklund transformations \cite{nah:09}.

\section{Lax pair}\label{sec:lax}

Nijhoff's method for obtaining a Lax pair for the Adler system \cite{n:02} is amenable to \eqref{eq:kdv}, and relies on the multidimensional consistency of the equation. If $\xu=x(m,n,l+1)$ denotes a shift in the lattice with parameter $r$ then we have
\begin{subequations}\label{eq:cac}
\begin{align}\label{eq:cac1}
Q_{pq}(x,\xt,\xh,\xth)&=0\\
\label{eq:cac2}
Q_{pr}(x,\xt,\xu,\xtu)&=0\\
\label{eq:cac3}
Q_{qr}(x,\xh,\xu,\xhu)&=0,
\end{align}
\end{subequations}
along with the group $D_4$ of square symmetries $Q_{pq}(x,\xt,\xh,\xth)=Q_{pq}(\xt,x,\xth,\xh)=Q_{qp}(x,\xh,\xt,\xth)$. Since \eqref{eq:kdv} is fractional linear in each variable we may solve \eqref{eq:cac2} and \eqref{eq:cac3} for $\xtu$ and $\xhu$ respectively
\begin{align}\label{eq:x3a}
\xtu&=\frac{x(\xt-\xu)+p^2-r^2}{\xt-\xu}\\
\label{eq:x3b}
\xhu&=\frac{x(\xh-\xu)+q^2-r^2}{\xh-\xu}.
\end{align}
The fact that \eqref{eq:cac2} and \eqref{eq:cac3} are both discrete Riccati equations for $\xu$ suggests the separation $\xu=f/g$, so $\xhu=\fh/\gh$ and $\xtu=\ft/\gt$. By defining $\bphi:=\left[\begin{array}{c}f\\g\end{array}\right]$ then \eqref{eq:x3a} and \eqref{eq:x3b} can be written in matrix form as
\begin{align}\label{eq:phi1}
\tilde{\bphi}&=\kappa_1 L\bphi\\
\label{eq:phi2}
\hat{\bphi}&=\kappa_2 M\bphi
\end{align}
where 
\begin{align}\label{eq:L}
L&=\left[\begin{array}{cc}-x&x\xt+p^2-r^2\\-1&\xt\end{array}\right]\\
\label{eq:M}
M&=\left[\begin{array}{cc}-x&x\xh+q^2-r^2\\-1&\xh\end{array}\right]
\end{align}
and $\kappa_1, \kappa_2$ are as yet undetermined separation functions. The determinantal condition \cite{n:02} on these functions is then
\begin{equation}
(\hat\kappa_1\kappa_2)^2(\det\hat L)(\det M)=(\kappa_1\tilde\kappa_2)^2(\det\tilde M)(\det L).
\end{equation}
Since $\det L=p^2-r^2$ and $\det M=q^2-r^2$ this allows for scalar (w.l.o.g. unity) values of the separation functions. One then finds that
\begin{equation}\label{eq:consistency}
\hat{\tilde\bphi}-\tilde{\hat\bphi}=Q_{pq}(x,\xt,\xh,\xth)\left[\begin{array}{cc}1&-(\xt+\xh)\\0&-1\end{array}\right]\bphi
\end{equation}
so that $\hat{\tilde\bphi}=\tilde{\hat\bphi}$ necessarily implies $x$ solves \eqref{eq:kdv}. 

The two systems of first-order difference equations \eqref{eq:phi1} and \eqref{eq:phi2} for $f$ and $g$ give rise to the second order difference equations for $g$
\begin{align}\label{eq:gtt}
\gtt-(\xtt-x)\gt+(p^2-r^2)g&=0 \\
\label{eq:ghh}
\ghh-(\xhh-x)\gh+(q^2-r^2)g&=0,
\end{align}
along with $\ft=\xt g-\gt$ and $\fh=\xh g-\gh$. Equations \eqref{eq:gtt} and \eqref{eq:ghh} serve as a Lax pair for the inverse scattering transform, since from \eqref{eq:consistency} we have $\gth-\ght=-Q_{pq}(x,\xh,\xt,\xth)g$. The lattice parameter $r$ acts as the spectral variable and the difference $\xtt-x$ acts as a potential. The difference equations in $f$ are auxiliary to the problem and are not required to be solved explicitly. 

\section{The Direct Scattering Problem}\label{sec:direct}

In this section we carry out the direct scattering procedure for equation \eqref{eq:gtt}, where we assume the solution $x$ of \eqref{eq:kdv} to be real. We first define $r:=iz$, so that $z$ now acts as the spectral variable, and alter the notation such that $g=g(m,n;z)$, $\gt=g(m+1,n;z)$, $\gh=g(m,n+1;z)$ and $x=x_{m,n}$, $\xt=x_{m+1,n}$, $\xh=x_{m,n+1}$.  

\subsection{Motivation}
One-soliton solutions to \eqref{eq:kdv} are given in \cite{av:04} and \cite{nah:09}. In \cite{av:04} the solution was obtained by taking a Backl\"und transformation of a linear seed solution and then using \eqref{eq:kdv} to obtain the $m$ and $n$ dependence, while in \cite{nah:09} the authors began with a Cauchy matrix structure and showed that it solved the homogeneous version of \eqref{eq:kdv}. The one-soliton solution is given by
\begin{equation}\label{eq:1solx}
x_{m,n}=pm+qn+C+\frac{2k}{A\rho_p^m\rho_q^n+1},
\end{equation}
where 
\begin{equation}\label{eq:rhop}
\rho_p=\frac{p+k}{p-k}, \; \; \; \rho_q=\frac{q+k}{q-k},
\end{equation}
and $C$ is constant. Thus for any $n$ the leading order behaviour of this solution as $m\rightarrow\pm\infty$ is
\[
x_{m,n}\sim pm+qn+const. \; {\rm as} \; m\rightarrow\pm\infty,
\]
where all other lower-order terms vanish exponentially. One finds the same result for the two-soliton solution given in Section \eqref{sec:12sol}. We therefore assume 
\begin{equation}
x_{m+2,n}-x_{m,n}=:2p+u_{m+1,n}
\end{equation}
where $u$ is a real-valued function satisfying $u\rightarrow0$ as $m\rightarrow\pm\infty$, independently of $n$. 

\subsection{Initial conditions}
As $n$ is arbitrary in the direct scattering problem we set $n=0$ and define $g(m;z)\equiv g(m,0;z)$ and $u_{m}\equiv u_{m,0}$. Since \eqref{eq:kdv} is invariant under the maps $p\mapsto-p$ and $q\mapsto-q$ we set $p>0$ and $q>0$ without loss of generality.  We also assume that
\begin{equation}\label{eq:upositivity}
2p+u_{m}>0 \; {\rm for}\;{\rm all} \; m,
\end{equation}
which is satisfied by \eqref{eq:1solx} provided we have $A>0$ and $p>k>0$, and by the two-soliton solution (with a similar restriction on parameters) given in Section \eqref{sec:12sol}. This assumption is sufficient\footnote{but not necessary. If $u_{m}=c(\delta_{m,0}+\delta_{m,1})$ then the case $c=-p(1+\sqrt2)$ gives
\[
a(z)=\frac{(z-i\frac p{\sqrt2})^2}{z(z+ip)}
\]
where the zeroes of $a(z)$, which is defined by \eqref{eq:ab}, are the discrete eigenvalues. For $c<-p(1+\sqrt2)$ however, the discrete eigenvalues are simple.} to prove that all discrete eigenvalues are simple (which was assumed without proof in \cite{bpps:01} and \cite{lp:07}), and as a consequence implies that all discrete eigenvalues are purely imaginary and lie within the interval $(0,ip)$. 

With the above assumptions the direct scattering problem is entirely governed by the second-order difference equation
\begin{equation}\label{eq:g}
g(m+2;z)-(2p+u_{m+1})g(m+1;z)+(p^2+z^2)g(m;z)=0.
\end{equation}

\subsection{Jost Solutions}
\begin{definition}\label{def:Jost}
The Jost solutions $(\varphi,\ovarphi)$ and $(\psi,\opsi)$ to \eqref{eq:g} are defined by the boundary conditions
\begin{subequations}\label{eq:jostboundary}
\begin{align}
\left\{\begin{array}{ll}
			\varphi&\sim(p-iz)^m \\ 
			\ovarphi&\sim(p+iz)^m
			\end{array}
			\right. 
{\rm as} \; m\rightarrow-\infty \\
\left\{\begin{array}{ll}
			\psi&\sim(p+iz)^m \\ 
			\opsi&\sim(p-iz)^m
			\end{array}\right. 
{\rm as} \; m\rightarrow+\infty,
\end{align}
\end{subequations}
where $(p\pm iz)^m$ solve \eqref{eq:g} as $u\rightarrow0$. 
\end{definition}

\subsection{Spectral Data}
Equation \eqref{eq:g} is invariant under the transformation $z\mapsto-z$, and by considering how the boundary conditions of the Jost solutions change under this mapping, uniqueness of the boundary value problem (see e.g. \cite{m:90}) implies that $\ovarphi(m;z)=\varphi(m;-z)$ and $\opsi(m;z)=\psi(m;-z)$. Since the general solution involves two linearly independent solutions we may write
\begin{equation}\label{eq:ab}
\psi=a\ovarphi+b\varphi,
\end{equation}
where $a=a(z)$ and $b=b(z)$, and thus $\opsi=a(-z)\varphi+b(-z)\ovarphi$. If $z$ is real then taking the complex conjugate of \eqref{eq:g} reveals $\ovarphi=\varphi^*$ and $\opsi=\psi^*$, and so $\psi=a\varphi^*+b\varphi$ on $\Im z=0$. If we define the  general solution to \eqref{eq:g} by the boundary conditions
\begin{subequations}\label{eq:RT}
\begin{align}
\frac{g(m;z)}{(p+iz)^m}&\sim1+R(z)\left(\frac{p-iz}{p+iz}\right)^m \; {\rm as} \; m\rightarrow-\infty \\
\frac{g(m;z)}{(p+iz)^m}&\sim T(z) \; {\rm as} \; m\rightarrow+\infty, 
\end{align}
\end{subequations}
where $R$ and $T$ are the transmission and reflection coefficients respectively, it follows that $g=\ovarphi+R\varphi$ and $g=T\psi=Ta\ovarphi+Tb\varphi$. Thus $R=\frac ba$ and $T=\frac1a$.

\begin{definition}\label{def:wronskian}

Let $g_1$ and $g_2$ be two solutions to \eqref{eq:g}. The discrete Wronskian $W_m(g_1,g_2;z)$ of $g_1(m;z)$ and $g_2(m;z)$ is then defined to be
\begin{equation}\label{eq:wronskian}
W_m(g_1,g_2;z)=\frac1{(p^2+z^2)^m}\left[g_1(m;z)g_2(m+1;z)-g_1(m+1;z)g_2(m;z)\right].
\end{equation}
\end{definition}

\begin{lemma}\label{lem:wronskian}

Suppose $g_1$ and $g_2$ are linearly independent nonzero solutions of \eqref{eq:g}. Then $W_m(g_1,g_2;z)$ is both nonzero and independent of $m$.
\begin{proof}
Since $g_1$ and $g_2$ both satisfy \eqref{eq:g} we have
\begin{align*}
&g_1(m+2;z)g_2(m+1;z)+(p^2+z^2)g_2(m+1;z)g_1(m;z) \\
=&(2p+u_{m+1})g_2(m+1;z)g_1(m+1;z) \\
=&g_2(m+2;z)g_1(m+1;z)+(p^2+z^2)g_1(m+1;z)g_2(m;z).
\end{align*}
Thus
\[
g_1(m;z)g_2(m+1;z)-g_1(m+1;z)g_2(m;z)=C_0(p^2+z^2)^m
\]
where $C_0$ is constant as required. If $W_m(g_1,g_2;z)=0$ then clearly $g_1$ and $g_2$ are linearly dependent.  
\end{proof}
\end{lemma}

\begin{proposition}\label{prop:ab}

For $\Im z=0$ the functions $a$ and $b$ defined by \eqref{eq:ab} satisfy
\begin{equation}
|a|^2-|b|^2=1.
\end{equation}
\begin{proof}
The linearity and anti-symmetry of the Wronskian imply that since $\psi=a\varphi^*+b\varphi$ along $\Im z=0$,
\[
W_m(\psi^*,\psi;z)=\bigl(|a(z)|^2-|b(z)|^2\bigr)W_m(\varphi,\varphi^*;z).
\]
As $W_m(\psi^*,\psi;z)$ and $W_m(\varphi,\varphi^*;z)$ are independent of $m$, they are equal to their boundary values at $m\rightarrow-\infty$ and $m\rightarrow+\infty$ respectively, which is $2iz$ in both cases. 
\end{proof}
\end{proposition}

\section{Analytic properties of the Jost solutions}\label{sec:Jost}

For the subsequent analysis it is convenient to make the following definition.

\begin{definition}\label{def:chiupsilon}
\begin{align}\label{eq:chi}
\chi(m;z)&:=\frac{\varphi(m;z)}{(p-iz)^m} \\
\label{eq:upsilon}
\Upsilon(m;z)&:=\frac{\psi(m;z)}{(p+iz)^m}.
\end{align}
Then $\chi\rightarrow1$ and $\Upsilon\rightarrow1$ as $m\rightarrow-\infty$ and $m\rightarrow+\infty$ respectively. $\overline{\chi}$ and $\overline{\Upsilon}$ are defined by $\overline{\chi}(m;z)=\chi(m;-z)$ and $\overline{\Upsilon}(m;z)=\Upsilon(m;-z)$. 
\end{definition}

\begin{lemma}\label{lem:jostsum}

For $\Im z\geq0$, $z\neq0$ the functions $\chi$ and $\Upsilon$ satisfy the following summation equations:
\begin{align}\label{eq:varphisum}
\chi(m;z)&=1+\frac1{2iz}\sum_{j=-\infty}^{m-1}\left[\left(\frac{p+iz}{p-iz}\right)^{m-j}-1\right]u_j\chi(j;z) \\
\label{eq:psisum}
\Upsilon(m;z)&=1+\frac1{2iz}\sum_{j=m+1}^{+\infty}\left[\left(\frac{p-iz}{p+iz}\right)^{m-j}-1\right]u_j\Upsilon(j;z) .
\end{align}
\begin{proof}
Equation \eqref{eq:g} for $\chi(m;z)$ gives
\[
(p-iz)\chi(m+2;z)-2p\chi(m+1;z)+(p+iz)\chi(m;z)=u_{m+1}\chi(m+1;z)
\]
which can be summed from an arbitrary $M_0$ to $m-1\geq M_0$ to give
\[
(p-iz)\bigl[\chi(m+1;z)-\chi(M_0+1;z)\bigr]-(p+iz)\bigl[\chi(m;z)-\chi(M_0;z)\bigr]=\sum_{j=M_0}^{m-1}u_{j+1}\chi(j+1;z). 
\]
Letting $M_{0}\rightarrow-\infty$ and incorporating the boundary behaviour of $\chi(m;z)$ gives
\[
\left(\frac{p-iz}{p+iz}\right)\chi(m+1;z)-\chi(m;z)=\frac{-2iz}{p+iz}+\frac1{p+iz}\sum_{j=-\infty}^mu_j\chi(j;z).
\]
We now multiply this equation by the summing factor $\left(\frac{p-iz}{p+iz}\right)^{m}$ and sum from an arbitrary $M_1$ to $m-1\geq M_1$. By letting $M_1\rightarrow-\infty$, incorporating the boundary behaviour of $\chi$ and noting that   $\left|\frac{p+iz}{p-iz}\right|\leq1$ for $\Im z\geq0$, this yields
\[
\chi(m;z)=1+\frac1{p+iz}\sum_{l=-\infty}^{m}\left(\frac{p+iz}{p-iz}\right)^{m-l}\sum_{j=-\infty}^lu_j\chi(j;z).
\]
Changing the order of summation then gives
\begin{align*}
\chi(m;z)&=1+\frac1{p+iz}\sum_{j=-\infty}^{m-1}u_j\chi(j;z)\sum_{l=j}^{m-1}\left(\frac{p+iz}{p-iz}\right)^{m-l} \\
&=1+\frac1{2iz}\sum_{j=-\infty}^{m-1}\left[\left(\frac{p+iz}{p-iz}\right)^{m-j}-1\right]u_j\chi(j;z).
\end{align*}
The proof of \eqref{eq:psisum} follows by a similar argument.
\end{proof}
\end{lemma}

\begin{lemma}\label{lem:jostsum0}
For $z=0$ the Jost solutions $\chi(m;0)$ and $\Upsilon(m;0)$ satisfy the following summation equations:
\begin{align}\label{eq:varphisum0}
\chi(m;0)&=1+\frac1p\sum_{j=-\infty}^{m-1}(m-j)u_j\chi(j;0) \\
\label{eq:psisum0}
\Upsilon(m;0)&=1+\frac1p\sum_{j=m+1}^{+\infty}(j-m)u_j\Upsilon(j;0).
\end{align}
\begin{proof}
\end{proof}
\end{lemma}

\begin{proposition}\label{prop:jostsol}

For $\Im z\geq0$, $z\neq0$ the summation equations \eqref{eq:varphisum} and \eqref{eq:psisum} have the following series solutions:
\begin{align}\label{eq:varphisol}
\chi(m;z)&=\sum_{k=0}^{+\infty}\frac{H_k(m;z)}{z^k} \\
\label{eq:psisol}
\Upsilon(m;z)&=\sum_{k=0}^{+\infty}\frac{J_k(m;z)}{z^k}
\end{align}
where
\begin{align}\label{eq:Hk}
H_0(m;z)=1&,\;H_{k+1}(m;z)=\frac1{2i}\sum_{j=-\infty}^{m-1}\left[\left(\frac{p+iz}{p-iz}\right)^{m-j}-1\right]u_jH_k(j;z) \\
\label{eq:Jk}
J_0(m;z)=1&,\;J_{k+1}(m;z)=\frac1{2i}\sum_{j=m+1}^{+\infty}\left[\left(\frac{p-iz}{p+iz}\right)^{m-j}-1\right]u_jJ_k(j;z).
\end{align}
\begin{proof}
Inserting \eqref{eq:varphisol} into the summation equation \eqref{eq:varphisum} gives
\begin{align*}
\chi(m;z)&=1+\frac1{2iz}\sum_{j=-\infty}^{m-1}\left[\left(\frac{p+iz}{p-iz}\right)^{m-j}-1\right]u_j\sum_{k=0}^{+\infty}\frac{H_k(j;z)}{z^k} \\
&=1+\sum_{k=0}^{+\infty}\frac1{z^{k+1}}\left(\frac1{2i}\sum_{j=-\infty}^{m-1}\left[\left(\frac{p+iz}{p-iz}\right)^{m-j}-1\right]u_jH_k(j;z)\right) \\
&=1+\sum_{k=0}^{+\infty}\frac{H_{k+1}(m;z)}{z^{k+1}}
\end{align*}
as required. The proof of \eqref{eq:psisol} is similar.
\end{proof}
\end{proposition}

\begin{proposition}\label{prop:jostsol0}
For $z=0$ the summation equations \eqref{eq:varphisum0} and \eqref{eq:psisum0} have the following series solutions:
\begin{align}\label{eq:varphisol0}
\chi(m;0)&=\sum_{k=0}^{+\infty}\frac{H_k^0(m)}{p^k} \\
\label{eq:psisol0}
\Upsilon(m;0)&=\sum_{k=0}^{+\infty}\frac{J_k^0(m)}{p^k}
\end{align}
where
\begin{align}\label{eq:Hk0}
H_0^0(m)=1&,\;H_{k+1}^0(m)=\sum_{j=-\infty}^{m-1}(m-j)u_jH_k^0(j) \\
\label{eq:Jk0}
J_0^0(m)=1&,\;J_{k+1}(m;z)=\sum_{j=m+1}^{+\infty}(j-m)u_jJ_k^0(j).
\end{align}
\begin{proof}
\end{proof}
\end{proposition}

The following theorems describe the analyticity properties of the Jost solutions in the $z$-plane. These results closely mirror those obtained for the direct scattering of the continuous Schr\"odinger equation, which can be found in the detailed analysis given in \cite{dt:79}.

\begin{theorem}\label{th:jostanalyticity}
Assume
\begin{equation}\label{eq:usummability}
\sum_{j=-\infty}^{+\infty}|u_j|(1+|j|)<\infty,
\end{equation}
Then for $\Im z\geq0$
\begin{align}\label{eq:chibound}
|\chi(m;z)-1|&\leq C_1 \; {\rm for}\; z\neq0 \\
\label{eq:chibound0}
|\chi(m;z)-1|&\leq C_2\left(1+\max\{m,0\}\right), \\
\end{align}
where $C_1$ and $C_2$ are constant. For all $m$, $\chi(m;z)$ is analytic in $\Im z>0$ and continuous in $\Im z\geq0$. For all $\Im z\geq0$ \eqref{eq:varphisol} converges absolutely in $m$ (and uniformly if $z\neq0$).

\begin{proof}
In the following all symbols $N_i$ refer to constants whose precise values are not required, but are used to obtain the required results. We first prove \eqref{eq:chibound}. For $\Im z\geq0$ we have $\left|\frac{p+iz}{p-iz}\right|\leq1$ and so the recursion relation \eqref{eq:Hk} for $\chi(m;z)$ can be upper-bounded by
\begin{equation}\label{eq:Hkupperbound}
|H_{k+1}(m;z)|\leq\sum_{j=-\infty}^{m-1}|u_j||H_k(j;z)|.
\end{equation}
\begin{lemma}\label{lem:Hkupperbound}
\[
|H_k(m;z)|\leq\frac{P(m-1)^k}{k!}
\]
where 
\[
P(m)=\sum_{j=-\infty}^m|u_j|.
\]
\begin{proof}
Clearly this holds for $k=0$. Equation \eqref{eq:Hkupperbound} then implies
\begin{align*}
|H_{k+1}(m;z)|&\leq\sum_{j=-\infty}^{m-1}|u_j|\frac{P(j-1)^k}{k!} \\
&=\frac1{k!}\sum_{j=-\infty}^{m-1}(P(j)-P(j-1))P(j-1)^k. 
\end{align*}
Summing by parts then gives
\begin{align*}
&\sum_{j=-\infty}^{m-1}(P(j)-P(j-1))P(j-1)^k=P(m-1)^{k+1}-\sum_{j=-\infty}^{m-1}(P(j)^k-P(j-1)^k)P(j) \\
&=P(m-1)^{k+1}-\sum_{j=-\infty}^{m-1}P(j)(P(j)-P(j-1))\times\left(\sum_{r=0}^{k-1}P(j)^{k-1-r}P(j-1)^r\right)  \\
&\leq P(m-1)^{k+1}-k\sum_{j=m+1}^{+\infty}(P(j)-P(j-1))P(j-1)^k,
\end{align*}
since $P(j-1)\leq P(j)$ for all $j$. This completes the inductive step.
\end{proof}
\end{lemma}

By \eqref{eq:varphisol} and Lemma \eqref{lem:Hkupperbound} we have
\[
\left|\chi(m;z)-1\right|\leq\sum_{k=1}^{+\infty}\frac{|H_k(m;z)|}{|z|^k}\leq\sum_{k=1}^{+\infty}\frac{P(m-1)^k}{|z|^kk!}\leq\frac{P(m)}{|z|}e^{\left(\frac{P(m)}{|z|}\right)}<N_1,
\]
which gives \eqref{eq:chibound}. Thus for any $z$ satisfying $\Im z\geq0$, $z\neq0$ the series solution for $\chi(m;z)$ converges absolutely and uniformly in $m$. An alternative upper bound for \eqref{eq:varphisum} is
\begin{align*}
|\chi(m;z)|&\leq1+\frac1{|p-iz|}\left|\sum_{j=-\infty}^{m-1}\left[\sum_{r=0}^{m-j-1}\left(\frac{p-iz}{p+iz}\right)^r\right]u_j\chi(j;z)\right| \\
&\leq1+\sigma\sum_{j=-\infty}^{m-1}(m-j)|u_j||\chi(j;z)|,
\end{align*}
where $\sigma=\max\{1,\frac1p\}$. Thus a majorant for either \eqref{eq:varphisum} or \eqref{eq:varphisum0} is
\begin{equation}\label{eq:chimajorant}
|\chi(m;z)|\leq1+\sigma\sum_{j=-\infty}^{m-1}(m-j)|u_j||\chi(j;z)|.
\end{equation}
Therefore for $\Im z\geq0$ 
\begin{equation}\label{eq:chiub}
|\chi(m;z)|\leq\sum_{k=0}^{+\infty}\bH_k(m)
\end{equation}
where
\[
\bH_{k+1}(m)=\sigma\sum_{j=-\infty}^{m-1}(m-j)u_j\bH_k(j).
\]
Following Lemma \eqref{lem:Hkupperbound} one can then show that
\[
\left|\bH_k(m)\right|\leq\frac{\sigma^kQ(m-1,m-1)^k}{k!},
\]
where $Q(m,L)=\sum_{j=-\infty}^L(m-j+1)|u_j|.$
Thus by \eqref{eq:chiub} we have
\begin{equation}\label{eq:sigmaR0}
\left|\chi(m;z)-1\right|\leq\sum_{k=1}^{+\infty}\frac{\sigma^kQ(m-1,m-1)^k}{k!}\leq\sigma Q(m,m)e^{\sigma Q(m,m)}.
\end{equation}
For $m\leq0$ equations \eqref{eq:chimajorant} and \eqref{eq:sigmaR0} give
\begin{align}
\left|\chi(m;z)-1\right|&\leq\sigma e^{\sigma Q(0,0)}\left[m\sum_{j=-\infty}^{m-1}|u_j|+\sum_{j=-\infty}^{m-1}(-j)|u_j|\right] \nonumber \\
&\leq\sigma e^{\sigma Q(0,0)}\left[\sum_{j=-\infty}^{-1}(-j)|u_j|\right] \nonumber \\
\label{eq:chimaxm<0}
&\leq N_2.
\end{align}
For $m>0$ equations \eqref{eq:chimajorant} and \eqref{eq:sigmaR0} give
\begin{align*}
|\chi(m;z)|&\leq1+\sigma\sum_{j=-\infty}^{m-1}(-j)|u_j||\chi(j;z)|+m\sigma\sum_{j=-\infty}^{m-1}|u_j||\chi(j;z)| \\
&\leq1+\sigma\sum_{j=-\infty}^{-1}(-j)|u_j||\chi(j;z)|+m\sigma\sum_{j=-\infty}^{m-1}|u_j||\chi(j;z)| \\
&\leq N_3+m\sigma\sum_{j=-\infty}^{m-1}|u_j||\chi(j;z)|.
\end{align*}
By writing $\chi(m;z)=N_3(1+m)\Xi_1(m;z)$ we see
\[
|\Xi_1(m;z)|\leq1+\sigma\sum_{j=-\infty}^{m-1}(1+|j|)|u_j||\Xi_1(j;z)|,
\]
which can be iterated to give
\begin{equation}\label{eq:chim>0}
|\Xi_1(m;z)|\leq\exp\left(\sigma\sum_{j=-\infty}^{m-1}(1+|j|)|u_j|\right) \; \Rightarrow \; |\chi(m;z)|\leq N_4(1+m).
\end{equation}
Thus for $m>0$, \eqref{eq:chimajorant}, \eqref{eq:sigmaR0} and \eqref{eq:chim>0} give
\begin{align}
|\chi(m;z)-1|&\leq\sigma\sum_{j=-\infty}^{-1}(-j)|u_j||\chi(j;z)|+m\sigma\sum_{j=-\infty}^{m-1}|u_j||\chi(j;z)| \nonumber \\
&\leq N_5+m\sigma N_4\sum_{j=-\infty}^{m-1}(1+|j|)|u_j|\nonumber \\
\label{eq:chimaxm>0}
&\leq N_6(1+m).
\end{align}
Combining this with the upper bound \eqref{eq:chimaxm<0} for $m\leq0$ gives \eqref{eq:chibound0}. This estimate is valid for all $m$ and all $z$ satisfying $\Im z\geq0$. For every $m$ the series solution \eqref{eq:varphisum} for $\chi(m;z)$ converges absolutely and uniformly in $z$ satisfying $\Im z\geq0$. Thus $\chi(m;z)$ is continuous in $z$ within $\Im z\geq0$. Since the iterates $H_k(m;z)$ are analytic functions of $z$ in $\Im z>0$, $\chi(m;z)$ is also analytic in this region. For $z=0$ the series solution for $\chi(m;0)$ converges absolutely in $m$ and uniformly for $m<m_0$.  

\end{proof}
\end{theorem}

\begin{remark}\label{rem:jostanalyticity}
By considering the series solution for $\Upsilon(m;z)$ one can similarly prove its existence and continuity in $\Im z\geq0$ and analyticity in $\Im z>0$, provided \eqref{eq:usummability} holds. One obtains similar estimates to \eqref{eq:chibound} and \eqref{eq:chibound0}.
\end{remark}

\begin{theorem}\label{thm:jostderivatives}
Assume that
\begin{equation}\label{eq:uL2}
\sum_{j=-\infty}^{+\infty}(1+j^2)|u_j|<\infty.
\end{equation}
Then for $\Im z\geq0$
\begin{equation}
\label{eq:chi'2}
|\chi'(m;z)|\leq C_3(1+m\max\{m,1\}),
\end{equation}
where $C_3$ is constant. For all $m$, $\chi'(m;z)$ exists and is continuous in $z$ for all $z$ in $\Im z\geq0$.
\begin{proof}
Again let $N_i$ denote constants as necessary. We rewrite \eqref{eq:varphisum} as 
\[
\chi(m;z)=1+\frac1{p-iz}\sum_{j=-\infty}^{m-1}\left(\sum_{r=0}^{m-j-1}\left(\frac{p+iz}{p-iz}\right)^r\right)u_j\chi(j;z),
\]
which agrees with \eqref{eq:varphisol0} at $z=0$ and is therefore valid everywhere in $\Im z\geq0$. Taking an upper bound of the derivative of this equation then shows
\begin{equation}\label{eq:chi'ub1}
|\chi'(m;z)|\leq\sigma^2\sum_{j=-\infty}^{m-1}(m-j)^2|u_j||\chi(j;z)|+\sigma\sum_{j=-\infty}^{m-1}(m-j)|u_j||\chi'(j;z)|.
\end{equation}
If $m\leq0$ then \eqref{eq:chibound0} implies
\[
\sum_{j=-\infty}^{m-1}(m-j)^2|u_j||\chi(j;z)|\leq\sum_{j=-\infty}^{m-1}j^2|u_j||\chi(j;z)|\leq N_1.
\]
If $m>0$ then \eqref{eq:chibound0} gives
\begin{align*}
&\sum_{j=-\infty}^{m-1}(m-j)^2|u_j||\chi(j;z)|\leq2\sum_{j=-\infty}^{m-1}m^2|u_j||\chi(j;z)|+2\sum_{j=-\infty}^{m-1}j^2|u_j||\chi(j;z)| \\
&\leq2\sum_{j=-\infty}^{-1}j^2|u_j||\chi(j;z)|+2m^2\sum_{j=1}^{m-1}|u_j||\chi(j;z)|+m^2\sum_{j=-\infty}^{m-1}|u_j||\chi(j;z)| \\
&\leq N_2+m^2N_3\sum_{j=-\infty}^{m-1}(1+|j|)|u_j| \\
&\leq N_4(1+m^2).
\end{align*}
Therefore \eqref{eq:chi'ub1} becomes
\begin{equation}\label{eq:chi'ub3}
|\chi'(m;z)|\leq\sigma^2N_4\left(1+m\max\{m,0\}\right)+\sigma\sum_{j=-\infty}^{m-1}(m-j)|u_j||\chi'(j;z)|
\end{equation}
which can be iterated to give
\[
|\chi'(m;z)|\leq\sigma^2N_4\left(1+m\max\{m,0\}\right)e^{\sigma Q(m-1,m-1)}.
\]
Inserting this into \eqref{eq:chi'ub3} then gives
\begin{align*}
|\chi'(m;z)|&\leq\sigma^2N_4\left(1+m\max\{m,0\}\right) \\
&+\sigma\sum_{j=-\infty}^{-1}(-j)|u_j||\chi'(j;z)|+\sigma m\sum_{j=-\infty}^{m-1}|u_j||\chi'(j;z)| \\
&\leq N_5\left(1+m\max\{m,0\}\right)+\sigma m\sum_{j=-\infty}^{m-1}|u_j||\chi'(j;z)|.
\end{align*}
For $m\leq0$
\[
|\chi'(m;z)|\leq N_5+\sigma m\sum_{j=-\infty}^{m-1}|u_j||\chi'(j;z)|,
\]
and so by defining $\chi'(m;z)=N_5(1+|m|)\Xi_2(m;z)$ we have
\[
|\Xi_2(m;z)|\leq1+\sigma\sum_{j=-\infty}^{m-1}(1+|j|)|u_j||\Xi_2(j;z)| \; \Rightarrow \; |\Xi_2(m;z)|\leq N_6,
\]
and so $|\chi'(m;z)|\leq N_7(1+|m|)$. For $m>0$ let $\chi'(m;z)=N_8(1+m^2)\Xi_3(m;z)$. Then
\[
|\Xi_3(m;z)|\leq1+\sigma\sum_{j=-\infty}^{m-1}(1+j^2)|u_j||\Xi_3(j;z)| \; \Rightarrow \; |\Xi_3(m;z)|\leq N_8, 
\]
and so $|\chi'(m;z)|\leq N_9(1+m^2)$. This proves \eqref{eq:chi'2}. For every $m$ the summation equation for $\chi'(m;z)$ converges absolutely and uniformly in $z$ satisfying $\Im z\geq0$, and absolutely in $m$. Therefore $\chi'(m;z)$ exists and is continuous for all $z$ in $\Im z\geq0$.  

\end{proof}
\end{theorem}

\begin{remark}
Using the summation equation for $\Upsilon$ one can similarly prove the existence and continuity of $\Upsilon'(m;z)$ in $\Im z\geq0$ provided \eqref{eq:usummability} holds. One obtains a similar estimate to \eqref{eq:chi'2}.
\end{remark}

\begin{corollary}\label{cor:jostbaranalyticity}
Assuming \eqref{eq:usummability} holds, the Jost solutions $\overline{\chi}$ and $\overline{\Upsilon}$ exist and are continuous in $\Im z\leq0$, and are analytic in $\Im z<0$. If \eqref{eq:uL2} also holds, then the derivatives of these functions also exist and are continuous in $\Im z\leq0$
\end{corollary}

\begin{remark}
The existence and continuity of the derivatives of the Jost solutions along $\Im z=0$ is not strictly required. Theorem \eqref{thm:jostderivatives} merely serves to illustrate under what restrictions on $u$ this condition will hold. Henceforth we only assume that $u$ satisfies \eqref{eq:usummability}.
\end{remark}

\section{Analyticity and Asymptotic Properties of $a(z)$ and $b(z)$}\label{sec:ab}

We now consider properties of the functions $a$ and $b$, which represent the transmission and reflection coefficients by the relations $R=\frac ba$ and $T=\frac1a$. In particular we look at their analyticity properties and asymptotic behaviour as $|z|\rightarrow+\infty$. We assume that \eqref{eq:usummability} holds.

\begin{proposition}\label{prop:abanalyticity}
The functions $a(z)$ and $b(z)$ defined by equation \eqref{eq:ab} have the following properties: 
\begin{enumerate}
\item[-] $a(z)$ is analytic in the region $\Im z>0$, and is continuous on $\Im z=0$, except possibly at $z=0$
\item[-] $b(z)$ is continuous on $\Im z=0$, except possibly at $z=0$.
\end{enumerate}
\begin{proof}
We have
\begin{align*}
W_m(\varphi,\psi;z)&=W_m(\varphi,a\ovarphi+b\varphi;z)=a(z)W_m(\varphi,\ovarphi;z)=2iza(z) \\
W_m(\psi,\ovarphi,;z)&=W_m(a\ovarphi+b\varphi,\ovarphi;z)=b(z)W_m(\varphi,\ovarphi;z)=2izb(z).
\end{align*}
Since $\varphi$ and $\psi$ are both analytic in $\Im z>0$ and continuous on $\Im z=0$, $a(z)$ also has this property, except possibly at $z=0$. The expression for $b(z)$ however is only valid on $\Im z=0$, where $\varphi$ and $\opsi$ are both defined. Thus $b(z)$ is continuous on $\Im z=0$, except possibly at $z=0$.
\end{proof}
\end{proposition}

\begin{proposition}\label{prop:absum}
For $z\neq0$ the functions $a(z)$ and $b(z)$ can be expressed as
\begin{align}\label{eq:asum}
a(z)&=1-\frac1{2iz}\sum_{j=-\infty}^{+\infty}u_j\Upsilon(j;z) \\
\label{eq:bsum}
b(z)&=\frac1{2iz}\sum_{j=-\infty}^{+\infty}u_j\Upsilon(j;z)\left(\frac{p+iz}{p-iz}\right)^{j}.
\end{align}
\begin{proof}
The summation equation \eqref{eq:varphisum} for $\chi$ can be expressed as
\begin{align}
\Upsilon(m;z)&=\left(1-\frac1{2iz}\sum_{j=m+1}^{+\infty}u_j\Upsilon(j;z)\right) \nonumber \\
\label{eq:varphisum2}
&+\left(\frac{p-iz}{p+iz}\right)^m\left(\frac1{2iz}\sum_{j=m+1}^{+\infty}u_j\Upsilon(j;z)\left(\frac{p+iz}{p-iz}\right)^{j}\right).
\end{align}
Comparing this with $\Upsilon(m;z)\sim a(z)+b(z)\left(\frac{p-iz}{p+iz}\right)^m$ as $m\rightarrow-\infty$ gives the desired result. 
\end{proof}
\end{proposition}

\begin{proposition}\label{prop:jostasymptotics}
\begin{align}\label{eq:varphiasymptotics}
\chi(m;z)&\sim1+O\left(\frac1{z}\right) \; {\rm as} \; |z|\rightarrow+\infty \;{\rm with}\; \Im z\geq0 \\
\label{eq:psiasymptotics}
\Upsilon(m;z)&\sim1+O\left(\frac1{z}\right) \; {\rm as} \; |z|\rightarrow+\infty \;{\rm with}\; \Im z\geq0.
\end{align}
\begin{proof}
This is clear from \eqref{eq:varphisol} and \eqref{eq:psisol}.
\end{proof}
\end{proposition}

\begin{corollary}\label{cor:jostbarasymptotics}
\begin{align}\label{eq:varphibarasymptotics}
\overline{\chi}(m;z)&\sim1+O\left(\frac1{z}\right) \; {\rm as} \; |z|\rightarrow+\infty \;{\rm with}\; \Im z\leq0 \\
\label{eq:psibarasymptotics}
\overline{\Upsilon}(m;z)&\sim1+O\left(\frac1{z}\right) \; {\rm as} \; |z|\rightarrow+\infty \;{\rm with}\; \Im z\leq0.
\end{align}
\end{corollary}

\begin{corollary}\label{cor:abasymptotics}
\begin{align}\label{eq:aasymptotics}
a(z)&\sim1-\frac1{2iz}\sum_{j=-\infty}^{+\infty}u_j+O\left(\frac1{z^2}\right) \; {\rm as} \; |z|\rightarrow+\infty \;{\rm with}\; \Im z\geq0, \\
\label{eq:basymptotics}
b(z)&\sim\frac1{2iz}\sum_{j=-\infty}^{+\infty}u_j\left(\frac{p+iz}{p-iz}\right)^j+O\left(\frac1{z^2}\right) \; {\rm as} \; |z|\rightarrow+\infty \;{\rm with}\; \Im z=0.
\end{align}
\begin{proof}
This follows directly from inserting the asymptotic behaviour \eqref{eq:psiasymptotics} of $\Upsilon$ into the expressions \eqref{eq:asum} and \eqref{eq:bsum} for $a$ and $b$. 

\end{proof}
\end{corollary}

\begin{theorem}
$a(z)$ has a finite number of zeroes $z_k$, $k=0,1,...,N$ in the open half-plane $\Im z>0$, and at each $z_k$ we have $\psi(m;z_k)=b_k\varphi(m;z_k)$, where $b_k=b(z_k)$. Moreover every zero is simple, lies on the imaginary axis and satisfies $|z_k|<p$. 
\begin{proof}
Proposition \eqref{prop:ab} implies $\Im z_k>0$, and since $a(z)\sim1+O\left(\frac1{z}\right)$ as $|z|\rightarrow\infty$ it follows that each $z_k$ must be finite in magnitude. It is also true that since $a(z)=\frac1{2iz}W_m(\varphi,\psi;z)$, Lemma \eqref{lem:wronskian} implies that $\psi(m;z_k)$ and $\varphi(m;z_k)$ are linearly dependent, so we write $\psi(m;z_k)=b_k\varphi(m;z_k)$, for some constant $b_k$. This implies
\[
\Upsilon(m;z_k)\sim b_k\left(\frac{p-iz_k}{p+iz_k}\right)^{m} \; {\rm as} \; m\rightarrow-\infty,
\]
so by equation \eqref{eq:varphisum2} we have 
\begin{align*}
2iz_k&=\sum_{j=-\infty}^{+\infty}u_j\Upsilon(m;z_k) \\
b_k&=\frac1{2iz_k}\sum_{j=-\infty}^{+\infty}u_j\Upsilon(m;z_k)\left(\frac{p+iz_k}{p-iz_k}\right)^{j},
\end{align*}
and thus $b_k=b(z_k)$. 

To show the zeroes of $a(z)$ are purely imaginary, consider
\[
\nu:=\frac{\psi}{(p^2+z^2)^{\frac m2}}.
\]
At $z=z_k$ we have
\begin{align*}
\nu&\sim b_k\left(\frac{p-iz_k}{p+iz_k}\right)^{\frac m2}\rightarrow0 \; {\rm as} \; m\rightarrow-\infty \\
\nu&\sim \left(\frac{p+iz_k}{p-iz_k}\right)^{\frac m2}\rightarrow0 \; {\rm as} \; m\rightarrow+\infty,
\end{align*}
and so $\sum_{j=-\infty}^{+\infty}\nu(j;z_k)$ exists. Equation \eqref{eq:g} for $\nu$ becomes
\begin{equation}\label{eq:nu}
(p^2+z^2)^{\frac12}\bigl(\nu(m+2;z)+\nu(m;z)\bigr)=(2p+u_{m+1})\nu(m+1;z), 
\end{equation}
and by fixing $z=z_k$, multiplying by $\nu^*(m+1;z_k)$ and summing each term from $m=-\infty$ to $+\infty$ we obtain
\[
(p^2+z_k^2)^{\frac12}\bigl(S+S^*)=\sum_{j=-\infty}^{+\infty}(2p+u_{j+1})|\nu(j,z_k)|^2,
\]
where $S=\sum_{j=-\infty}^{+\infty}\nu(j+1;z_k)\nu^*(j;z_k^*)$. Since we assume $(2p+u_{m+1})>0$ for all $m$ it follows that the right hand side of this equation cannot vanish, and so $z_k^2$ must be real. This in turn implies that $\psi(m;z_k)$ is a real-valued function. Hence by squaring \eqref{eq:nu} and fixing $z=z_k$ we see
\begin{equation}\label{eq:eigensign}
(p^2+z_k^2)\left[\nu(m+2;z_k)+\nu(m;z_k)\right]^2=\left[(2p+u_{m+1})\nu(m+1;z_k)\right]^2>0,
\end{equation}
which implies that $p^2-|z_k|^2>0$.

To show that the zeroes of $a(z)$ are simple we note that since $a(z)=\frac1{2iz}W_m(\varphi,\psi;z)$, it follows that
\begin{equation}\label{eq:a'}
2iz_ka'(z_k)=\frac d{dz}\bigl. W(\varphi,\psi;z)\bigr|_{z=z_k}=W(\varphi',\psi;z_k)-W(\psi',\varphi;z_k).
\end{equation}
Taking a $z$-derivative of \eqref{eq:g} shows that $\varphi'$ satisfies
\begin{align*}
&\varphi'(m+2;z)+(p^2+z^2)\varphi'(m;z)+2z\varphi(m;z)=(2p+u_{m+1})\varphi'(m+1;z) \\
\Rightarrow &\varphi'(m+2;z)\psi(m+1;z)+(p^2+z^2)\psi(m+1;z)\varphi'(m;z)+2z\psi(m+1;z)\varphi(m;z) \\
&=(2p+u_{m+1})\varphi'(m+1;z)\psi(m+1;z).
\end{align*}
Comparing this with equation \eqref{eq:g} for $\psi$ shows
\begin{align*}
&\varphi'(m+2;z)\psi(m+1;z)+(p^2+z^2)\psi(m+1;z)\varphi'(m;z)+2z\psi(m+1;z)\varphi(m;z) \\
&=\psi(m+2;z)\varphi'(m+1;z)+(p^2+z^2)\varphi'(m+1;z)\psi(m;z),
\end{align*}
which after dividing through by $(p^2+z^2)^{m+1}$ and summing from $m_0\leq m$ to $m$ reveals
\[
\bigl.W(\varphi',\psi;z)\bigr|^{m}_{m_0}=2z\sum_{j=m_0}^{m-1}\frac{\psi(j+1;z)\varphi(j;z)}{(p^2+z^2)^{j+1}}.
\]
The same procedure can then be carried out with $\varphi$ and $\psi$ interchanged, except choosing to sum from $m$ to $m_1\geq m$:
\[
\bigl.W(\psi',\varphi;z)\bigr|^{m_1}_{m}=2z\sum_{j=m}^{m_1}\frac{\varphi(j+1;z)\psi(j;z)}{(p^2+z^2)^{j+1}}.
\]
Now fix $z=z_k$. Since $\psi(m;z_k)=b_k\varphi(m;z_k)$ it follows that
\begin{align*}
W(\varphi',\psi;z_k)&\sim ib_k\left(\frac{p-iz_k}{p+iz_k}\right)^m\rightarrow0 \; {\rm as} \; m\rightarrow-\infty \\
W(\psi',\varphi;z_k)&\sim \frac{-i}{b_k}\left(\frac{p+iz_k}{p-iz_k}\right)^m\rightarrow0 \; {\rm as} \; m\rightarrow+\infty,
\end{align*}
and so by taking $m_0=-\infty$, $m_1=+\infty$, equation \eqref{eq:a'} shows
\[
a'(z_k)=\frac{-i}{b_k}\sum_{j=-\infty}^{+\infty}\frac{\psi(j+1;z_k)\psi(j;z_k)}{(p^2+z_k^2)^{j+1}}.
\]
From \eqref{eq:g} however we have
\[
\psi(m+2;z_k)\psi(m+1;z_k)+(p^2+z_k^2)\psi(m+1;z_k)\psi(m;z_k)=(2p+u_{m+1})\psi^2(m+1;z_k),
\]
which implies
\begin{equation}\label{eq:a'2}
a'(z_k)=\frac{-i}{b_k}\sum_{j=-\infty}^{+\infty}\frac{(2p+u_{j+1})\psi^2(j+1;z_k)}{2(p^2+z_k^2)^{j+1}}.
\end{equation}
As $\psi(m;z_k)$ is a real-valued function this final sum, and hence $a'(z_k)$, is nonzero as required.

Finally since $a(z)$ is analytic in $\Im z>0$, it must have isolated zeroes in a finite region along the positive imaginary axis. The only way that $a(z)$ could have an infinite number of zeroes in this region is if they formed a limiting sequence which accumulated at $z=0$. We show that this is not possible. Assume that there exists a sequence of zeroes of $a(z)$, $\{z_k\}$, which all lie on the imaginary axis and accumulate at $z=0$: $\lim_{k\rightarrow\infty}z_k=0$. Then at each $z_k$ we have 
\[
b(z_k)=\frac{\psi(m;z_k)}{\varphi(m;z_k)},
\]
and so 
\[
\lim_{k\rightarrow\infty}|b(z_k)-b(0)|=\lim_{k\rightarrow\infty}\left|\frac{\psi(m;z_k)}{\varphi(m;z_k)}-\frac{\psi(m;0)}{\varphi(m;0)}\right|=0
\]
since the Jost solutions are continuous at $z=0$. At $z=0$ however we have $\varphi(m;0)=\ovarphi(m;0)$ and so $\psi(m;0)=(a(0)+b(0))\varphi(m;0)$. But
\begin{align*}
&\lim_{k\rightarrow\infty}b(z_k)=\lim_{k\rightarrow\infty}\frac{\psi(m;z_k)}{\varphi(m;z_k)} \\
&\Rightarrow b(0)=a(0)+b(0) \\
&\Rightarrow a(0)=0,
\end{align*}
which contradicts the fact that $|a(0)|^2=1+|b(0)|^2\geq1$. Thus $a$ has a finite number of zeroes in $\Im z>0$. 
\end{proof}
\end{theorem}

\section{Evolution of the Scattering Data in the $n$-direction}\label{sec:n}

Equation \eqref{eq:ghh} governs the behaviour of solutions in the $n$-direction:
\begin{equation}\label{eq:gn}
g(m,n+2;z)-(x_{m,n+2}-x_{m,n})g(m,n+1;z)+(q^2+z^2)g(m,n;z)=0.
\end{equation}
The one- and two-soliton solutions behave like $x_{m,n}\sim pm+qn+const.$ as $m\rightarrow\pm\infty$ independently on $n$, where all lower-order terms vanish exponentially. We therefore assume that $x_{m,n+2}-x_{m,n}\rightarrow2q$ as $m\rightarrow\pm\infty$ for all $n$.

\begin{definition}
The $n$-dependent Jost solutions $\varphi^{(n)}$ and $\psi^{(n)}$ to equation \eqref{eq:gn} are defined to be
\begin{align}\label{eq:varphin}
\varphi^{(n)}&=(q-iz)^n\varphi \\
\label{eq:psin}
\psi^{(n)}&=(q+iz)^n\psi.
\end{align}
$\ovarphi^{(n)}$ and $\opsi^{(n)}$ are defined by $\ovarphi^{(n)}(m,n;z)=\varphi^{(n)}(m,n;-z)$ and $\opsi^{(n)}(m,n;z)=\psi^{(n)}(m,n;-z)$. 
\end{definition}

The $n$ evolution of $\varphi^{(n)}$ reveals that $\varphi$ satisfies
\begin{equation}\label{eq:varphineq}
(q-iz)\varphi(m,n+2;z)-(x_{m,n+2}-x_{m,n})\varphi(m,n+1;z)+(q+iz)\varphi(m,n;z)=0.
\end{equation}
Note that \eqref{eq:gn} does not allow for the $n$-independent boundary conditions of $\varphi$ and $\psi$, but these are consistent with \eqref{eq:varphineq}. Since $\psi(m,n;z)=a(n;z)\ovarphi(m,n;z)+b(n;z)\varphi(m,n;z)$ holds for all $n$, we have
\begin{equation}\label{eq:abn}
\psi^{(n)}(m,n;z)=a(n;z)\ovarphi^{(n)}(m,n;z)+b(n;z)\left(\frac{q+iz}{q-iz}\right)^n\psi^{(n)}(m,n;z),
\end{equation}
which agrees with \eqref{eq:ab} on $n=0$ if we adopt the notation $a(z)\equiv a(0;z)$ and $b(z)\equiv b(0;z)$. 

\begin{theorem}\label{thm:abn}
The function $a(n;z)$ is independent of $n$: 
\begin{equation}\label{eq:an}
a(n;z)=a(0;z)\equiv a(z).
\end{equation}
The function $b(n;z)$ is given by
\begin{equation}\label{eq:bn}
b(n;z)=b(0;z)\left(\frac{q-iz}{q+iz}\right)^n\equiv b(z)\left(\frac{q-iz}{q+iz}\right)^n.
\end{equation}
\begin{proof}
In the limit $m\rightarrow-\infty$ we have $\varphi^{(n)}(m,n;z)\sim(p-iz)^m(q-iz)^n$ and $\ovarphi^{(n)}(m,n;z)\sim(p+iz)^m(q+iz)^n$, and so in this limit $\varphi^{(n)}$ and $\ovarphi^{(n)}$ are linearly independent solutions of \eqref{eq:gn}. Thus we may write
\[
\psi^{(n)}(m,n;z)\sim C_1(m;z)\ovarphi^{(n)}(m,n;z)+C_2(m;z)\varphi^{(n)}(m,n;z) \; {\rm as} \; m\rightarrow-\infty,
\]
where $C_1$ and $C_2$ are independent of $n$. Comparing this with \eqref{eq:abn} gives the desired result.
\end{proof}
\end{theorem}

\section{Inverse Scattering}\label{sec:inverse}

We now proceed to reconstruct the potential $u$, and ultimately the solution $x$ of \eqref{eq:kdv}. We rewrite equation \eqref{eq:ab} as
\begin{equation}\label{eq:jump}
\frac{\Upsilon(m;z)}{a(z)}-\overline{\chi}(m;z)=R(z)\chi(m;z)\left(\frac{p-iz}{p+iz}\right)^m,
\end{equation}
where $R(z)=\frac{b(z)}{a(z)}$ is the reflection coefficient. This equation defines a jump condition  along $\Im z=0$ between $\frac{\Upsilon}{a}$, which is meromorphic in the open half-plane $\Im z>0$, and $\overline{\chi}$, which is analytic in the open half plane $\Im z<0$. Coupled with the knowledge of the behaviour of these functions for large $|z|$, namely
\begin{align}\label{eq:RHboundary1}
\frac{\Upsilon(m;z)}{a(z)}&\sim\left(1+O\left(\frac1{z}\right)\right) \; {\rm as} \; |z|\rightarrow+\infty, \Im z\geq0 \\
\label{eq:RHboundary2}
\overline{\chi}(m;z)&\sim\left(1+O\left(\frac1{z}\right)\right) \; {\rm as} \; |z|\rightarrow+\infty, \Im z\leq0,
\end{align}
this becomes an example of the classical Riemann-Hilbert problem. To solve this problem we consider the Cauchy integral defined along the real $z$-axis:
\begin{equation}\label{eq:I}
I:=\frac1{2\pi i}\int_{-\infty}^{+\infty}\frac{\Upsilon(m;\zeta)}{a(\zeta)(\zeta+z)}d\zeta \; \mathrm{with} \; \Im z>0.
\end{equation}
We first evaluate this integral by considering a semi-circular contour in the upper-half plane $\Im\zeta\geq0$. Let $\Gamma_+$ denote the arc $\zeta=Me^{i\theta}$ with $0\leq\theta\leq\pi$ and $M\gg1$. Along this contour we have
\begin{align*}
\frac1{2\pi i}\int_{\Gamma_+}\frac{\Upsilon(m;\zeta)}{a(\zeta)(\zeta+z)}d\zeta&=\frac1{2\pi i}\int_0^{\pi}\frac{iMe^{i\theta}}{Me^{i\theta}+z}d\theta\left(1+O\left(\frac1M\right)\right) \\
&=\frac1{2\pi}\int_0^{\pi}d\theta\left(1+O\left(\frac1M\right)\right) \\
&=\frac12\left(1+O\left(\frac1M\right)\right).
\end{align*}
Since $a$ has a finite number of simple zeroes at $z=z_k$ in $\Im z>0$ and $\zeta+z$ cannot vanish in this region, the residue theorem gives
\begin{align}
I&=\sum_{k=1}^{N}\lim_{\zeta\rightarrow z_k}\frac{\Upsilon(m;\zeta)(\zeta-z_k)}{a(\zeta)(\zeta+z)}-\frac12 \nonumber \\
\label{eq:I1}
&=\sum_{k=1}^{N}\frac{\epsilon_k\chi(m;z_k)}{(z+z_k)}\left(\frac{p-iz_k}{p+iz_k}\right)^m-\frac12
\end{align}
where
\begin{equation}\label{eq:epsilonk}
\epsilon_k=b_k\lim_{\zeta\rightarrow z_k}\frac{(\zeta-z_k)}{a(\zeta)}=\frac{b_k}{a'(z_k)},
\end{equation}
and we have used the fact that $\Upsilon(m;z_k)=b_k\chi(m;z_k)\left(\frac{p-iz_k}{p+iz_k}\right)^m$. Note that by Theorem \eqref{thm:abn} $\epsilon_k$ depends on $n$. We may also calculate the Cauchy integral \eqref{eq:I} by using the jump condition \eqref{eq:jump},
\begin{equation}\label{eq:I2}
I:=\frac1{2\pi i}\int_{-\infty}^{+\infty}\frac{\overline{\chi}(m;\zeta)}{(\zeta+z)}d\zeta+\frac1{2\pi i}\int_{-\infty}^{+\infty}\frac{R(\zeta)\chi(m;\zeta)}{(\zeta+z)}\left(\frac{p-i\zeta}{p+i\zeta}\right)^md\zeta,
\end{equation}
and considering a semi-circular contour in the lower half-plane $\Im\zeta\leq0$.  Let $\Gamma_-$ denote the arc $\zeta=Me^{i\theta}$ with $-\pi\leq\theta\leq0$ and $M\gg1$. Then
\begin{align*}
\frac1{2\pi i}\int_{\Gamma_-}\frac{\overline{\chi}(m;\zeta)}{(\zeta+z)}d\zeta&=\frac1{2\pi i}\int_{-\pi}^0\frac{iMe^{i\theta}}{Me^{i\theta}+z}d\theta\left(1+O\left(\frac1M\right)\right) \\
&=\frac12\left(1+O\left(\frac1M\right)\right),
\end{align*}
and since $\zeta+z$ will vanish in $\Im\zeta<0$, the residue theorem gives
\begin{align*}
\frac1{2\pi i}\int_{-\infty}^{+\infty}\frac{\overline{\chi}(m;\zeta)}{(\zeta+z)}d\zeta&=-\lim_{\zeta\rightarrow-z}\frac{\overline{\chi}(m;\zeta)(\zeta+z)}{(\zeta+z)}+\frac12 \\
&=-\overline{\chi}(m;-z)+\frac12 \\
&=-\chi(m;z)+\frac12.
\end{align*}
Thus \eqref{eq:I2} gives
\[
I=-\chi(m;z)+\frac12+\frac1{2\pi i}\int_{-\infty}^{+\infty}\frac{R(\zeta)\chi(m;\zeta)}{(\zeta+z)}\left(\frac{p-i\zeta}{p+i\zeta}\right)^md\zeta,
\]
and by combining this with \eqref{eq:I1} to eliminate $I$ we obtain
\begin{equation}\label{eq:RHresult}
\chi(m;z)=1-\sum_{k=1}^{N}\frac{\epsilon_k\chi(m;z_k)}{(z+z_k)}\left(\frac{p-iz_k}{p+iz_k}\right)^m+\frac1{2\pi i}\int_{-\infty}^{+\infty}\frac{R(\zeta)\chi(m;\zeta)}{(\zeta+z)}\left(\frac{p-i\zeta}{p+i\zeta}\right)^md\zeta.
\end{equation}

For the following results we introduce the convenient notation
\[
\lambda:=\left(\frac{p-iz}{p+iz}\right).
\]

\begin{lemma}\label{lemma:orthonomality}
The function $\lambda^m$ obeys the orthonormality condition
\begin{equation}\label{eq:weight}
\int_{-\infty}^{+\infty}\lambda^{n-m}w(z)dz=\delta_{nm},
\end{equation}
where $w(z)=\frac1\pi\frac{p}{p^2+z^2}$.
\begin{proof}
This follows directly from using the substitution $\theta=\tan^{-1}\left(\frac{z}{p}\right)$. 
\end{proof}
\end{lemma}

\begin{proposition}\label{prop:T}
$\chi$ is expressible as the discrete integral transform of $\lambda^{r-m}$:
\begin{align}
\chi(m;z)&=1+{\mathcal T}[\lambda^{r-m}], \nonumber \\
\label{eq:T}
&:=1+\frac{2p}{p-iz}\sum_{r=-\infty}^{m}K(m,r)\lambda^{r-m},
\end{align}
where the kernel $K(m,r)$ is independent of $z$. 
\begin{proof}
Substituting \eqref{eq:T} into \eqref{eq:g} and using $p-iz=\frac{2p\lambda}{1+\lambda}$, $p+iz=\frac{2p}{1+\lambda}$  gives
\begin{align*}
&2p\sum_{r=-\infty}^{m+1}K(m+2,r+1)\lambda^{r-m}-(2p+u_{m+1})\sum_{r=-\infty}^{m}K(m+1,r+1)\lambda^{r-m} \\
-&(2p+u_{m+1})\sum_{r=-\infty}^{m+1}K(m+1,r)\lambda^{r-m}+2p\sum_{r=-\infty}^{m}K(m,r)\lambda^{r-m} \\
&+\lim_{r\rightarrow-\infty}\bigl[\bigl(2pK(m+2,r)-(2p+u_{m+1})K(m+1,r)\bigr)\lambda^{r-m-1}\bigr]=\lambda u_{m+1}.
\end{align*}
For such a $K(m,r)$ to exist for $r\leq m$ we therefore require that it satisfy
\begin{equation} \label{eq:K}
2p\left[K(m+2,r+1)+K(m,r)\right]-(2p+u_{m+1})\left[K(m+1,r+1)+K(m+1,r)\right]=0,
\end{equation}
subject to the boundary conditions
\begin{subequations}\label{eq:KBC}
\begin{align}\label{eq:KBC1}
2pK(m+2,m+2)-(2p+u_{m+1})K(m+1,m+1)&=u_{m+1} \\
\label{eq:KBC2}
\lim_{r\rightarrow-\infty}K(m,r)&=0.
\end{align}
\end{subequations}
Note that we assume that $K(m,r)\rightarrow0$ fast enough so that $\sum_{r=-\infty}^{m}|K(m,r)|$ exists, since this is sufficient for \eqref{eq:T} to exist along $\Im z=0$. We now show that a solution to \eqref{eq:K} with boundary conditions \eqref{eq:KBC} exists and is unique for all $r\leq m$. By defining $F_0(m):=K(m,m)$ the boundary condition \eqref{eq:KBC1} gives an inhomogeneous first-order difference equation for $F_0$,
\[
2pF_0(m+2)-(2p+u_{m+1})F_0(m+1)=u_{m+1},
\]
which can be solved to give the one-parameter family of solutions
\[
F_0(m)=-1+A_0\prod_{i=-\infty}^{m-1}\left(1+\frac{u_i}{2p}\right).
\]
The convergence of this product follows from
\[
\prod_{i=-\infty}^{+\infty}\left|1+\frac{u_i}{2p}\right|\leq\prod_{i=-\infty}^{+\infty}\left(1+\frac{|u_i|}{2p}\right)\leq\exp\left(\frac1{2p}\sum_{i=-\infty}^{+\infty}|u_{i}|\right)<\infty.
\]
The second boundary condition \eqref{eq:KBC2} then imposes $A_0=1$ and so $F_0$ is determined uniquely. Now set $r=m$ in \eqref{eq:K} and define $F_1(m)=K(m,m-1)$. The equation then reads
\[
F_1(m+2)-(1+\frac{u_{m+1}}{2p})F_1(m+1)=(1+\frac{u_{m+1}}{2p})F_0(m+1)-F_0(m),
\]
which is an inhomogeneous first-order difference equation for $F_1$ in terms of $F_0$. This can be solved using a summing factor and then the boundary condition \eqref{eq:KBC2} will determine $F_1$ uniquely. Now set $r=m-1$ in \eqref{eq:K} and define $F_2(m):=K(m,m-2)$. This gives an inhomogeneous first-order equation for $F_2$ in terms of $F_1$, which can be solved uniquely in the above fashion. Thus by induction we have a unique solution $K(m,m-\alpha)$ for all $m$, where $\alpha$ is any non-negative integer.
\end{proof}
\end{proposition}

\begin{proposition}\label{prop:Tinverse}
For $L\leq m$, the kernel $K(m,L)$ defined by \eqref{eq:T} can be expressed as
\begin{equation}\label{eq:Tinverse}
K(m,L)=\frac1{2\pi}\int_{-\infty}^{+\infty}{\mathcal T}[\lambda^{r-m}]\lambda^{m-L}\frac{dz}{p+iz}.
\end{equation}
\begin{proof}
\begin{align*}
&\frac1{2\pi}\int_{-\infty}^{+\infty}{\mathcal T}[\lambda^{r-m}]\lambda^{m-L}\frac{dz}{p+iz} \\
=&\frac1{2\pi}\int_{-\infty}^{+\infty}\left(\frac{2p}{p-iz}\sum_{r=-\infty}^{m}K(m,r)\lambda^{r-m}\right)\lambda^{m-L}\frac{dz}{p+iz} \\
=&\sum_{r=-\infty}^{m}K(m,r)\frac1{\pi}\int_{-\infty}^{+\infty}\lambda^{r-L}\frac{p}{p^2+z^2}dz \\
=&\sum_{r=-\infty}^{m}K(m,r)\delta_{rL} \\
=&K(m,L).
\end{align*}
\end{proof}
\end{proposition}

By inserting \eqref{eq:T} into \eqref{eq:RHresult} and simplifying one then obtains
\begin{align}
&\frac{2p}{p-iz}\sum_{r=-\infty}^{m}K(m,r)\lambda^{r-m}+\sum_{k=1}^{N}\frac{\epsilon_k}{z+z_k}\left(\frac{p-iz_k}{p+iz_k}\right)^m \nonumber \\
+&\sum_{r=-\infty}^{m}K(m,r)\sum_{k=1}^{N}\frac{2p\epsilon_k}{(z+z_k)(p-iz_k)}\left(\frac{p-iz_k}{p+iz_k}\right)^r-\frac1{2\pi i}\int_{-\infty}^{+\infty}\frac{R(\zeta)}{\zeta+z}\left(\frac{p-i\zeta}{p+i\zeta}\right)^md\zeta \nonumber \\
\label{eq:RHK}
-&\frac1{2\pi i}\sum_{r=-\infty}^{m}K(m,r)\int_{-\infty}^{+\infty}\frac{2pR(\zeta)}{(\zeta+z)(p-i\zeta)}\left(\frac{p-i\zeta}{p+i\zeta}\right)^rd\zeta=0.
\end{align}
We now perform the integral in \eqref{eq:Tinverse} in order to recover $K$. To do this we multiply the equation by $\frac{\lambda^{m-L}}{2\pi(p+iz)}$, where $L\leq m$, and as we have assumed $\Im z>0$ we integrate in $z$ from $-\infty$ to $+\infty$ along a path $P_+$ just above the real axis. 

\begin{lemma}\label{lem:Tinverseintegrals}
For $L\leq m$, $z_k$ purely imaginary with $|z_k|>0$ and $\Im\zeta=0$ we have the following results:
\begin{align}\label{eq:Tint1}
&\int_{-\infty}^{+\infty}\frac{\lambda^{m-L}}{(z+z_k)(p+iz)}dz=\frac{-2\pi i}{p-iz_k}\left(\frac{p-iz_k}{p+iz_k}\right)^{L-m} \\
\label{eq:Tint2}
&\int_{P_+}\frac{\lambda^{m-L}}{(z+\zeta)(p+iz)}dz=\frac{-2\pi i}{(p-i\zeta)}\left(\frac{p-i\zeta}{p+i\zeta}\right)^{L-m}
\end{align}
\begin{proof}
This follows by considering the poles of the integrand and using the residue theorem appropriately.
\end{proof}
\end{lemma}

By Lemma \eqref{lem:Tinverseintegrals}, taking the aforementioned integral of equation \eqref{eq:RHK} gives a discrete Gel'fand-Levitan integral equation for $K$:
\begin{equation}\label{eq:GL}
K(m,L)+B(L)+\sum_{r=-\infty}^{m}K(m,r)(B(r-m+L)+B(r-m+L-1))=0,
\end{equation}
where
\begin{equation}\label{eq:B1}
B(T):=\sum_{k=1}^{N}\frac{-i\epsilon_k}{(p-iz_k)}\left(\frac{p-iz_k}{p+iz_k}\right)^T+\frac1{2\pi}\int_{-\infty}^{+\infty}\frac{R(\zeta)}{(p-i\zeta)}\left(\frac{p-i\zeta}{p+i\zeta}\right)^{T}d\zeta.
\end{equation}

\begin{proposition}\label{prop:GL}
Suppose that for fixed $m$, $|B(L)|\leq|B(m)|$ for $L\leq m$, and $\sum_{-\infty}^m|B(r)|$ exists. Then the discrete Gel'fand-Levitan equation \eqref{eq:GL} has a unique solution $K(m,L)$ for $L\leq m$.
\begin{proof}
We first prove existence. If we express $K$ as
\begin{equation}\label{eq:kseries}
K(m,L)=\sum_{j=0}^{+\infty}H_j(m,L),
\end{equation}
then this will solve \eqref{eq:GL} if 
\begin{align*}
H_{j+1}&=-\sum_{r=-\infty}^mH_j(m,r)(B(r-m+L)+B(r-m+L-1)) \\
H_0&=-B(L).
\end{align*}
For $L\leq m$ the recursion relation can be upper-bounded by
\[
|H_{j+1}(m,L)|\leq2\sum_{r=-\infty}^m|B(r)||H_j(m,r)|,
\]
and so emulating the inductive process in Lemma \eqref{lem:Hkupperbound} one obtains
\[
H_j(m,L)\leq\frac{2^j\left(\sum_{r=-\infty}^m|B(r)|\right)^j}{j!}.
\]
Thus \eqref{eq:kseries} converges absolutely and uniformly for $L\leq m$. If $K_1$ and $K_2$ both solve \eqref{eq:GL} then $\beta:=K_1-K_2$ satisfies
\[
|\beta(m,L)|\leq\sum_{r=-\infty}^m2|B(r)||\beta(m,r)|.
\]
If $|\beta(m,L)|\leq N$ for $L\leq m$ then by the above inductive argument we obtain
\[
|\beta(m,L)|\leq N\frac{2^j\left(\sum_{r=-\infty}^m|B(r)|\right)^j}{j!}
\]
for any $j\geq0$, thus $\beta=0$.
\end{proof}
\end{proposition}

The function $B$ is dependent on the quantities
\begin{equation}\label{eq:S}
S:=\left\{R(n;\zeta) \; {\rm for} \; \Im\zeta=0 \; ; \; z_k \; ; \; \epsilon_k(n)\right\}
\end{equation}
which collectively comprise the scattering data. The entire inverse scattering process can therefore be summarised as follows: Given a potential $u_{m}$ defined on $n=0$ that satisfies $2p+u_{m}>0$ for all $m$ and 
\[
\sum_{j=-\infty}^{+\infty}(1+|j|)|u_j|<\infty,
\]
one may construct the the Jost solutions $\varphi(m;z)$ and $\psi(m;z)$ using equation \eqref{eq:g}. This leads to the knowledge of the functions $a$ and $b$ and thus to the $n$-dependent scattering data through equations \eqref{eq:an} and \eqref{eq:bn}. The kernel $K(m,L)$ can then be calculated via equation \eqref{eq:GL}, and the $n$-dependent potential reconstructed from the boundary condition \eqref{eq:KBC}:
\begin{equation}\label{eq:u!}
u_{m,n}=2p\left[\frac{1+K(m+1,m+1)}{1+K(m,m)}-1\right].
\end{equation}
The solution of \eqref{eq:kdv} can then be found by first writing $x_{m,n}=pm+qn+C+f(m,n)$, where $C$ is arbitrary and $f\rightarrow0$ as $|m|\rightarrow+\infty$, and then integrating $f(m+2,n)-f(m,n)=u_{m+1,n}$. This gives
\begin{equation}\label{eq:x!}
x_{m,n}=pm+qn+C+\sum_{r\geq1, r\;{\rm odd}}u_{m-r,n}.
\end{equation}

\section{One- and Two-Soliton Examples}\label{sec:12sol}

The potential $u_{m+1}$ along $n=0$ for the one-soliton solution \eqref{eq:1solx} is
\begin{equation}\label{eq:1solu}
u_{m+1}=-2k\frac{A\rho_p^m(\rho_p^2-1)}{\left(A\rho_p^{m+2}+1\right)\left(A\rho_p^m+1\right)}.
\end{equation}
Solving \eqref{eq:g} for $g(m;z)$ gives the Jost solutions as
\begin{subequations}\label{eq:1soljost}
\begin{align}
\frac{\varphi(m;z)}{(p-iz)^m}&=\left(\frac{A\rho_p^ma(z)+1}{A\rho_p^m+1}\right) \\
\frac{\psi(m;z)}{(p+iz)^m}&=\left(\frac{A\rho_p^m+a(z)}{A\rho_p^m+1}\right),
\end{align}
\end{subequations}
where
\begin{subequations}
\begin{align}
a(z)&=\frac{z-ik}{z+ik} \\
b(z)&=0.
\end{align}
\end{subequations}
At $z_1=ik$ we have $\psi(m;z_1)=A\varphi(m;z_1)$ so $b_1(n;z)=A\rho_q^{n}$ and
\[
\epsilon_1=2ikA\rho_q^{n}.
\]
Since $b(n;z)\equiv0$ we have $R(z)\equiv0$, and
\[
B(T)=\frac{2kA}{p+k}\rho_p^{T}\rho_q^{n} \\
\]
The discrete Gel'fand-Levitan equation \eqref{eq:GL} for $K(m,L)$ then becomes
\[
K(m,L)+\frac{2kA}{p+k}\rho_p^L\rho_q^n+\frac{4pkA}{(p+k)^2}\rho_q^n\sum_{r=-\infty}^{m}K(m,r)\rho_p^{r-m+L}=0.
\]
It is natural to assume the form $K(m,L)=Q(m)\rho_p^L\rho_q^n$, yielding
\[
Q(m)=\frac{A(\rho_p^{-1}-1)}{1+A\rho_p^m\rho_q^n} \; \Rightarrow \; 1+K(m,m)=\frac{1+A\rho_p^{m-1}\rho_q^n}{1+A\rho_p^m\rho_q^n}.
\]
Thus by \eqref{eq:u!}
\begin{align*}
u_{m,n}&=2p\frac{(1+A\rho_p^m\rho_q^n)^2-(1+A\rho_p^{m-1}\rho_q^n)(1+A\rho_p^{m+1}\rho_q^n)}{(1+A\rho_p^{m-1}\rho_q^n)(1+A\rho_p^{m+1}\rho_q^n)} \\
&=-2p\frac{A\rho_p^{m-1}\rho_q^n(\rho_p-1)^2}{(1+A\rho_p^{m-1}\rho_q^n)(1+A\rho_p^{m+1}\rho_q^n)} \\
&=\frac{2k}{A\rho_p^{m+1}\rho_q^n}-\frac{2k}{A\rho_p^{m-1}\rho_q^n},
\end{align*}
where we have used the fact that $2p=2k\left(\frac{\rho_p+1}{\rho_p-1}\right)$. Finally by \eqref{eq:x!} we have
\[
x_{m,n}=pm+qn+C+\frac{2k}{1+A\rho_p^m\rho_q^n}.
\]

Two-soliton solutions to \eqref{eq:kdv} are given in \cite{av:04} and \cite{nah:09}, and are again derived by Backl\"und transformation and Cauchy matrix approach respectively. The solution along $n=0$ takes the form
\[
x_{m,0}=pm+C+\frac{2(k_1+k_2)+2k_1A_1\rho_{p1}^m+2k_2A_2\rho_{p2}^m}{1+A_1\rho_{p1}^m+A_2\rho_{p2}^m+\left(\frac{k_1-k_2}{k_1+k_2}\right)^2A_1A_2\rho_{p1}^m\rho_{p2}^m},
\]
where
\[
\rho_{p_j}:=\left(\frac{p+k_j}{p-k_j}\right),\;\rho_{q_j}:=\left(\frac{q+k_j}{q-k_j}\right).
\]
From this one may calculate $u_{m+1}$, and then solve \eqref{eq:g} to determine the Jost solutions as
\begin{subequations}\label{eq:jost2sol}
\begin{align}
\frac{\varphi(m;z)}{(p-iz)^m}&=\frac{1+\left(\frac{z-ik_1}{z+ik_1}\right)A_1\rho_{p1}^m+\left(\frac{z-ik_2}{z+ik_2}\right)A_2\rho_{p2}^m+\left(\frac{k_1-k_2}{k_2+k_1}\right)^2a(z)A_1A_2\rho_{p1}^m\rho_{p2}^m}{1+A_1\rho_{p1}^m+A_2\rho_{p2}^m+\left(\frac{k_1-k_2}{k_1+k_2}\right)^2A_1A_2\rho_{p1}^m\rho_{p2}^m} \\
\frac{\psi(m;z)}{(p+iz)^m}&=\frac{a(z)+\left(\frac{z-ik_2}{z+ik_2}\right)A_1\rho_{p1}^m+\left(\frac{z-ik_1}{z+ik_1}\right)A_2\rho_{p2}^m+\left(\frac{k_1-k_2}{k_2+k_1}\right)^2A_1A_2\rho_{p1}^m\rho_{p2}^m}{1+A_1\rho_{p1}^m+A_2\rho_{p2}^m+\left(\frac{k_1-k_2}{k_1+k_2}\right)^2A_1A_2\rho_{p1}^m\rho_{p2}^m},
\end{align}
where
\begin{align*}
a(z)&=\frac{(z-ik_1)(z-ik_2)}{(z+ik_1)(z+ik_2)} \\
b(z)&=0.
\end{align*}
\end{subequations}
Thus again this corresponds to a reflectionless potential, but we now have two discrete eigenvalues $z_1=ik_1$ and $z_2=ik_2$. At each of these we have
\[
\epsilon_1(n)=2ik_1A_1\rho_{q1}^n \; \; \; \epsilon_2(n)=2ik_2A_2\rho_{q2}^n,
\]
and the solution of the Gel'fand levitan is
\begin{align*}
1+&K(m,m)= \\
&\frac{1+A_1\rho_{p1}^{m-1}\rho_{q1}^n+A_2\rho_{p2}^{m-1}\rho_{q2}^n+\left(\frac{k_1-k_2}{k_1+k_2}\right)^2A_1A_2\rho_{p1}^{m-1}\rho_{p2}^{m-1}\rho_{q1}^n\rho_{q2}^n}{1+A_1\rho_{p1}^m\rho_{q1}^n+A_2\rho_{p2}^m\rho_{q2}^n+\left(\frac{k_1-k_2}{k_1+k_2}\right)^2A_1A_2\rho_{p1}^m\rho_{p2}^m\rho_{q1}^n\rho_{q2}^n}.
\end{align*}
After simplifying the expression \eqref{eq:u!} for $u_{m,n}$ and using \eqref{eq:x!} one then obtains
\[
x_{m,n}=pm+qn+C+\frac{2(k_1+k_2)+2k_2A_1\rho_{p1}^m\rho_{q1}^n+2k_1A_2\rho_{p2}^m\rho_{q2}^n}{1+A_1\rho_{p1}^m\rho_{q1}^n+A_2\rho_{p2}^m\rho_{q2}^n+\left(\frac{k_1-k_2}{k_1+k_2}\right)^2A_1A_2\rho_{p1}^m\rho_{q1}^n\rho_{p2}^m\rho_{q2}^n},
\]
which is (after redefining constants as necessary) the two-soliton solution given in \cite{av:04} and \cite{nah:09}.

\section{Aribitrary Reflectionless Potential}\label{sec:Nsol}

We now consider an arbitrary reflectionless potential, that is one that satisfies $b(z)=0$ on $\Im z=0$.\footnote{One can show that if \eqref{eq:uL2} holds, this implies that 
\[
a(z)=\prod_{\iota=1}^{N}\left(\frac{z-ik_\iota}{z+ik_\iota}\right).
\]
} We write the $N$ discrete eigenvalues as $z_j=ik_j$, and make the further assumption that $0<k_j<q$ for all $j$. The Gel'fand-Levitan equation reads
\begin{equation}\label{eq:GLNsol}
K(m,L)+\sum_{j=1}^{N}\frac{-i\epsilon_j}{(p+k_j)}\rho_{p_j}^L+\sum_{j=1}^{N}\frac{-2ip\epsilon_j}{(p+k_j)^2}\sum_{r=-\infty}^{m}K(m,r)\rho_{p_j}^{r-m+L}=0.
\end{equation}
From equation \eqref{eq:a'2} we have
\begin{equation}\label{eq:iepsNsol}
-i\epsilon_j=\left(\sum_{r=-\infty}^{+\infty}\frac{(2p+u_r)\varphi^2(r;z_k)}{2(p^2+z_k^2)^r}\right)^{-1}>0,
\end{equation}
and so we set $-i\epsilon_j=:c_j\rho_j^o\rho_{q_j}^n$, where $c_j$ and $\rho_j^o$ are constants satisfying $c_j\rho_j^o>0$. To solve \eqref{eq:GLNsol} we assume the form
\begin{equation}\label{eq:KQ}
K(m,L)=-\sum_{\nu=1}^{N}Q_\nu(m)\frac1{p+k_\nu}\rho_\nu^o\rho_{p_\nu}^L\rho_{q_\nu}^n,
\end{equation}
which gives a system of $N$ equations of the form
\begin{equation}\label{eq:GLQ}
Q_j(m)+\sum_{\nu=1}^NQ_\nu(m)\frac{\rho_\nu^o\rho_{p_\nu}^{m}\rho_{q_\nu}^{n}c_j}{k_\nu+k_j}=c_j.
\end{equation}
 By defining
\begin{align}
{\bf Q}(m)^T&:=\left[Q_1(m),...,Q_N(m)\right] \\
\label{eq:reflect2}
{\bf c}^T&:=[c_1,c_2,...,c_N] \\
\label{eq:reflect3}
M_{\nu j}&:=\frac{\rho_\nu^o\rho_{p_\nu}^{m}\rho_{q_\nu}^{n}c_j}{k_\nu+k_j}
\end{align}
we may express \eqref{eq:GLQ} in matrix form as
\[
{\bf Q}(m)^T(I+M)={\bf c}^T.
\]
Since $M$ may be written as the Cauchy matrix $A_{\nu j}=\frac1{k_\nu+k_j}$ multiplied on the left by the diagonal matrix of elements $\rho_\nu^o\rho_{p_\nu}^{m}\rho_{q_\nu}^{n}$, and on the right by the diagonal matrix of elements $c_j$, the determinant formula for the Cauchy matrix $A_{\nu j}$ gives
\[
\det M=\left(\prod_{\nu=1}^{N}\frac{c_\nu\rho_\nu^o\rho_{p_\nu}^{m}\rho_{q_\nu}^{n}}{2k_\nu}\right)\prod_{\nu<j}\left(\frac{k_\nu-k_j}{k_\nu+k_j}\right)^2>0.
\]
Thus every term in the expansion for $\det(I+M)$ will be positive, and so $I+M$ is invertible. Thus by \eqref{eq:KQ}
\begin{equation}\label{eq:KNsol}
K(m+1,m+1)=-{\bf c}^T(I+\widetilde M)^{-1}(pI-\kappa)^{-1}{\bf r},
\end{equation}
where $\; \widetilde{} \;$ indicates the shift $m\mapsto m+1$, $\kappa_{\nu j}$ is the diagonal matrix with entries $\kappa_{\nu j}=\delta_{\nu j}k_j$ and ${\bf r}$ is the vector with entries
\begin{equation}\label{eq:reflect4}
r_j=\rho_j^o\rho_{p_j}^m\rho_{q_j}^n.
\end{equation}

\begin{theorem}\label{thm:Nsol}

Any solution $x_{m,n}$ of \eqref{eq:kdv} that gives rise to a reflectionless potential $u_{m+1}$ satisfying \eqref{eq:upositivity} can be expressed as
\begin{equation}\label{eq:xNsol}
x_{m,n}=pm+qn+C-{\bf c}^T(I+M)^{-1}{\bf r},
\end{equation}
where $M$ is defined by \eqref{eq:reflect3} with positive constants $c_j$, $C$ is arbitrary and ${\bf r}$ is defined by \eqref{eq:reflect4}.

\begin{proof}
We show that \eqref{eq:xNsol} satisfies \eqref{eq:u!} with $K(m+1,m+1)$ defined by \eqref{eq:KNsol}. Using the identities
\begin{align*}
(I+\widetilde M)(pI+\kappa)-(pI+\kappa)(I+M)&={\bf \widetilde r}{\bf c}^T \\
(pI-\kappa)(I+\widetilde M)-(I+M)(pI-\kappa)&={\bf r}{\bf c}^T,
\end{align*}
we have
\begin{align*}
&K(m+1,m+1)u_{m+1,n}=-{\bf c}^T(I+M)^{-1}{\bf r}{\bf c}^T(I+\widetilde M)^{-1}(pI-\kappa)^{-1}{\bf r} \\
&+{\bf c}^T(I+\widetilde{\widetilde M})^{-1}{\bf \widetilde{\widetilde r}}{\bf c}^T(I+\widetilde M)^{-1}(pI-\kappa)^{-1}{\bf r} \\
=&-{\bf c}^T(I+M)^{-1}\bigl[(pI-\kappa)(I+\widetilde M)-(I+M)(pI-\kappa)\bigr](I+\widetilde M)^{-1}(pI-\kappa)^{-1}{\bf r} \\
&+{\bf c}^T(I+\widetilde{\widetilde M})^{-1}\bigl[(I+\widetilde{\widetilde M})(pI+\kappa)-(pI+\kappa)(I+\widetilde M)\bigr](I+\widetilde M)^{-1}(pI-\kappa)^{-1}{\bf r} \\
=-&{\bf c}^T(I+M)^{-1}{\bf r}+2p{\bf c}^T(I+\widetilde M)^{-1}(pI-\kappa)^{-1}{\bf r}-{\bf c}^T(I+\widetilde{\widetilde M})^{-1}{\bf \widetilde{\widetilde r}},
\end{align*}
where we have used the fact that $(pI+\kappa)(pI-\kappa)^{-1}{\bf r}={\bf \widetilde r}$. Thus
\begin{align*}
(1+K(m+1,m+1))u_{m+1,n}&=-2p{\bf c}^T(I+\widetilde{\widetilde M})^{-1}(pI-\kappa)^{-1}{\bf \widetilde{\widetilde r}} \\
&+2p{\bf c}^T(I+\widetilde M)^{-1}(pI-\kappa)^{-1}{\bf \widetilde r} \\
&=2p(K(m+2,m+2)-K(m+1,m+1))
\end{align*}
as required.
\end{proof}
\end{theorem}

\begin{remark}
The $N$-soliton solution given in \cite{nah:09} is
\begin{align*}
x_{m,n}=A&\left(pm+qn+C-{\bf c}^T(I+M)^{-1}{\bf r}\right) \\
&+B(-1)^{m+n}\left(pm+qn+D-{\bf c}^T(I+M)^{-1}{\bf r}\right),
\end{align*}
where $A^2-B^2=1$ and $C$ and $D$ are arbitrary constants. Thus the solution \eqref{eq:xNsol} is the particular case of this solutions with $A=1$ and $B=0$. This is due to the fact that we have assumed $x\sim pm+qn+C$ as $m\rightarrow\pm\infty$. This is not restrictive however as the full general solution can be obtained by noticing that if $x_{m,n}$ solves \eqref{eq:kdv} then so does 
\[
w_{m,n}:=Ax_{m,n}+B(-1)^{m+n}(x_{m,n}+const.),
\]
provided that $A^2-B^2=1$. Thus we find that our arbitrary reflectionless solution \eqref{eq:xNsol} agrees exactly with that given in \cite{nah:09}, the only difference being that here we have restricted the parameters such that \eqref{eq:upositivity} holds. 
\end{remark}

\section{Conclusion}

In this paper we have developed an inverse scattering transform for the LKdV \eqref{eq:kdv}. Using the 3D consistency of the equation we generated a Lax pair following \cite{n:02} and used this to determine the discrete Schr\"odinger equation \eqref{eq:g} and its direct scattering problem. This was solved in section \eqref{sec:direct}, in which results were proved rigorously and precise estimates were obtained. The discrete ``time'' evolution of the scattering data was calculated in Section \eqref{sec:n} using the second Lax equation, and in Section \eqref{sec:inverse} the inverse problem, which was posed along the real axis of the spectral parameter, was solved. Rather than giving the solution in terms of the Jost solutions we emulated the continuous Riemann-Hilbert approach and derived a discrete Gel'fand-Levitan equation, whose solution is related to the potential by \eqref{eq:u!}. The solution of the lattice equation \eqref{eq:kdv} is then given by \eqref{eq:x!}. In Section \eqref{sec:12sol} the one-soliton and  two-soliton solutions were given as examples and these were found to correspond to reflectionless potentials with one and two discrete eigenvalues respectively, as expected. The arbitrary reflectionless potential case with $N$ discrete eigenvalues was then shown in Section \eqref{sec:Nsol} to correspond exactly to the $N$-soliton solution given in \cite{nah:09}, except with certain parameter restrictions. 

The inverse scattering transform works for all potentials $u_{m,n}$ satisfying the summability condition \eqref{eq:usummability}, which is the discrete analogue of the integrability condition in the continuous case, and the positivity condition \eqref{eq:upositivity}. This positivity condition is sufficient to prove that the poles of the tranmission coefficient are simple, however it is not necessary as was shown in a counterexample. A more precise restriction on the potential is therefore desirable, and will be addressed in future investigations.

\section{Acknowledgements}
This research was funded by the Australian Research Council grant DP0985615.

\end{document}